\documentstyle[epsfig]{aipproc}

\def\mdcrit{{\dot m}_{\rm crit}}
\def\dm{{\dot m}}

\begin{document}

\title{Geometry of accretion in Soft X-ray Transients}

\author{P.T. \.{Z}ycki$^*$, C. Done$^*$ and D.A. Smith$^{\dagger}$}
\address{$^*$Department of Physics, University of Durham, 
Durham DH1 3LE, England \\
$^{\dagger}$Department of Physics and Astronomy, University of Leicester,
       Leicester LE1 7RH, England}

\maketitle

\begin{abstract}

We present results of modelling {\it Ginga}\/ data of a number 
of Soft X-ray Transient sources performed with the aim of constraining
the geometry of accretion during various stages of the sources' evolution.
Assuming a generic geometry of a central, extended X-ray source with an 
external, optically thick accretion disk,
we use consistent models of X-ray reprocessing to constrain the inner radius
of the accretion disk and its ionization state. We show that the evolution
of GS~1124-68 (Nova Muscae 1991) and GS~2000+25 agree qualitatively
with the recent 
ideas linking the high/low state transition with changing radius of the
optically thick disk,
but the case of GS~2023+33 (V404 Cyg) poses a problem for any model.

\end{abstract}

\section*{Introduction}

Black hole binary systems
give one of the most direct ways in which to study the
physics of accretion disks since there is no surface boundary layer or strong 
central
magnetic field to disrupt the flow and the  parameters of the systems
are often very well known.
Additionally, many of these systems (the Soft X--ray Transients,
hereafter SXT's) show dramatic outbursts where the luminosity rises rapidly 
to close to the Eddington limit, and
then declines over a period of months, giving a clear 
sequence of spectra as a function of mass accretion rate. 
Usually, the outburst spectrum is dominated by a soft component of
 temperature $\sim 1$ keV, with
or without a power law tail, while the later stages of the
decline show much harder power law spectra, extending out to 100--200 keV
\cite{tash96}.
The same bimodal spectral states
are seen in the persistent black hole candidates, showing
that they are a general outcome of an accretion flow.

The  ``standard'''' accretion disk model \cite{shsu73} (hereafter SS)
gives temperatures of order 1~keV for galactic black hole 
candidates  (GBHC) at high accretion rates but it is 
unable to explain the presence of the power law tail to high energies, 
possibly indicating the presence of an additional, optically thin 
phase of accretion flow.

Recently there has been much excitement about the possibility that 
advection dominated accretion flows (ADAF; see \cite{nara97} for review)
may explain the hard X--ray data.
These flows can exist only below a critical accretion rate, $\dm\le \mdcrit$
and, by assumption, their radiative efficiency is very low.
Since ADAFs are hot and optically thin, they lack any strong source of 
soft seed photons for Comptonization which means 
that the resulting X--ray spectra are hard.
Such flows were proposed to explain
the hard and very faint X--ray spectra seen from SXT 
in quiescence \cite{nmcy96}, and then extended 
in \cite{esin97} (hereafter EMN) to cover the whole range of luminosity 
seen in SXT's. As $\dm$ increases to $\dm\sim \mdcrit$ the radiative 
efficiency of the flow
increases, but above $\dm=\mdcrit$
the advective flow collapses into an SS disk. 
This change from a hot ADAF to a cool SS disk is proposed to be the origin 
of the hard/soft  spectral transition seen in GBHC (see Fig.~1 in EMN). 

The changing geometry of EMN's model
is {\it testable\/}  from the X--ray spectral data.  
Wherever hard X--rays illuminate optically thick material
they can be reflected back through electron scattering. 
The reflected spectrum
is harder than the intrinsic spectrum, with photo--electric edge
features and the associated fluorescence lines imprinted on it,
most prominently from  iron \cite{gefa91}.
It is  a function of the ionization state and elemental abundances of the
reflecting gas\cite{ross93}. 

Furthermore, the sharp spectral features can be
broadened and smeared  by the Doppler effect due to
high orbital velocities and gravitational redshift 
if the reflection is from an accretion disk orbiting a black hole 
\cite{frsw89}.
Therefore, the reprocessed component gives us a strong 
diagnostic of the geometry:
the amount of reflection indicates the solid angle subtended by the
optically thick material, while the relativistic smearing 
reveals its velocity field.
In particular, 
reflected features can only be seen if there is optically
thick material that subtends a substantial solid angle to the X--ray source,
unlike the ADAF--based geometry proposed for the low and quiescence 
state SXT's. 

\section*{Data modelling and the geometry}

First, we used {\it Ginga}\/ data of  Nova Muscae 1991
to look for the effects of reprocessing, 
since it is the source used by EMN to illustrate their model, 
and it shows the full sequence of spectral evolution from Very High State
(VHS) to Low State (LS).
We fit these data with a self--consistent model of the reflected
continuum and Fe K$\alpha$ line, including relativistic
smearing \cite{zds97}.

\begin{figure} 
\centerline{
\epsfig{file=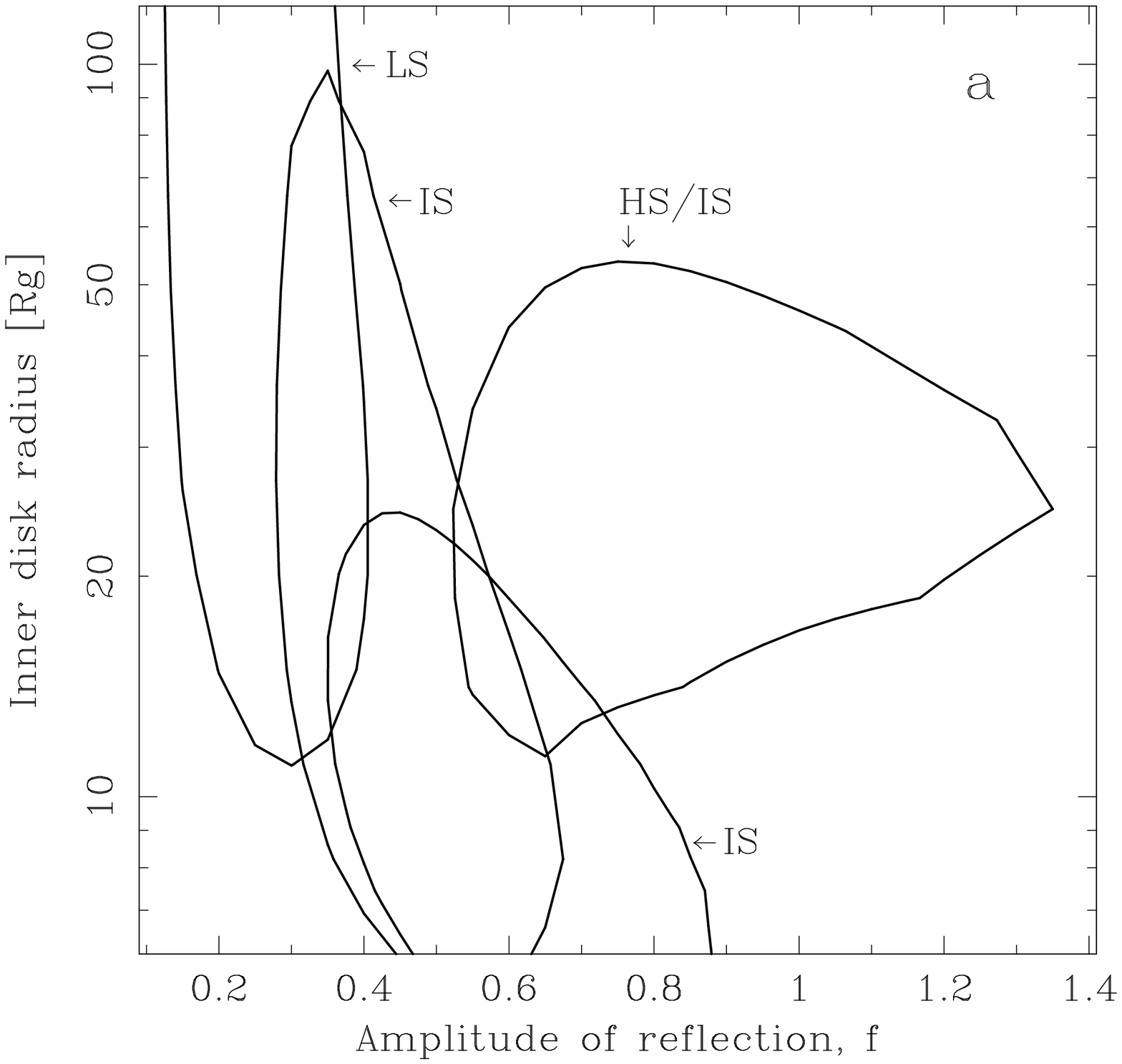,width=2.2in,height=2.2in }
\hspace{40pt}
\epsfig{file=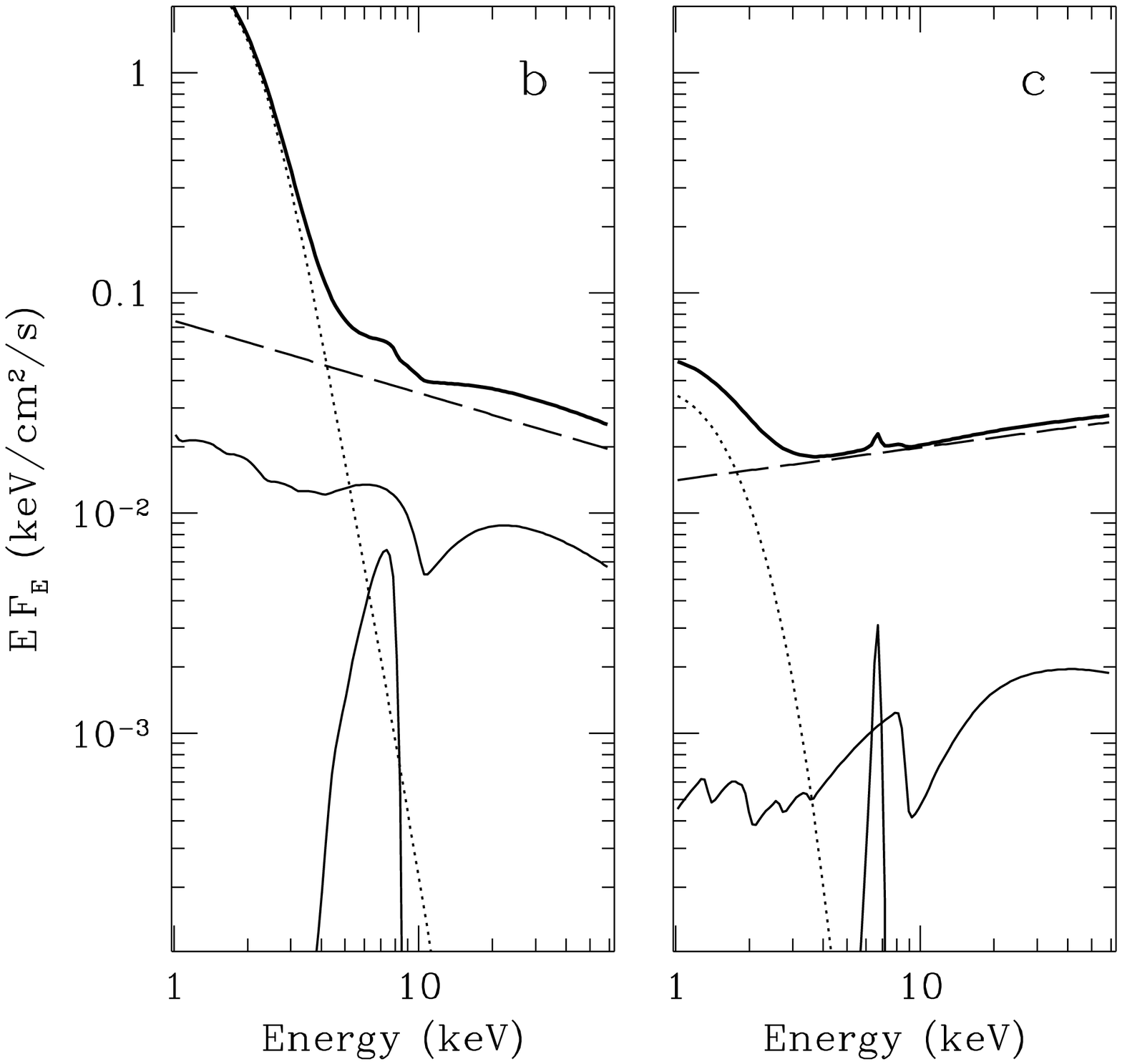,width=2.2in,height=2.2in}
}
\vspace{10pt}
\caption{
(a) --  confidence contours as functions of amplitude of reflection and
inner radius of the reflecting disk, for a sequence of spectra of Nova Muscae
1991; (b), (c) -- best fit model spectra for GS~2000+25: (b) -- soft/high 
state, (c) -- hard/low state. 
}\label{zds:fig1}
\end{figure}
 
In the VHS the reflector is highly ionised and strongly
relativistically smeared. The power law component is too weak to examine
for reflected features in the
High State (HS), but at the start of the Intermediate State (IS) 
the reflector is still
highly ionised and strongly smeared. It seems probable that
during all of this time the disk extends down to the last stable orbit, 
and that it is ionised by the strong
soft component. As source fades through the IS and LS,
importance of the soft component declines, the ionization of the reflector
drops, the power law hardens by $\Delta\Gamma\sim 0.4$, and the solid angle
subtended by the reflector decreases.
Figure~\ref{zds:fig1}a
shows the derived confidence contours for the innermost radius of the
disk (from constraints on relativistic smearing) and amount of reflection
($f=1$ corresponds to the amount of reflection expected from a reprocessor 
which covers $2\pi$ solid angle) during the IS and LS.

\begin{figure} 
\centerline{
\epsfig{file=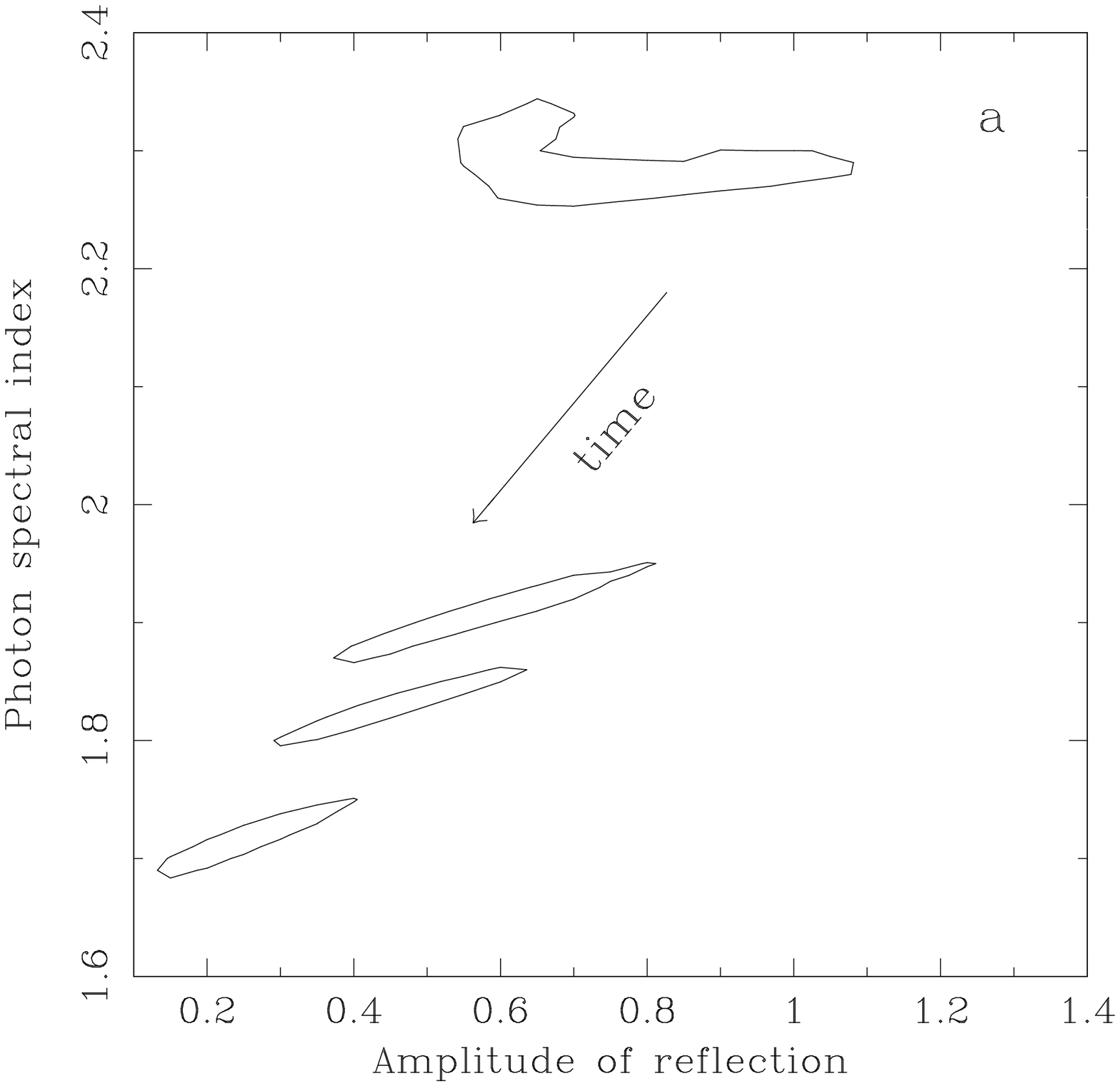,width=2.2in,height=2.2in}
\hspace{40pt}
\epsfig{file=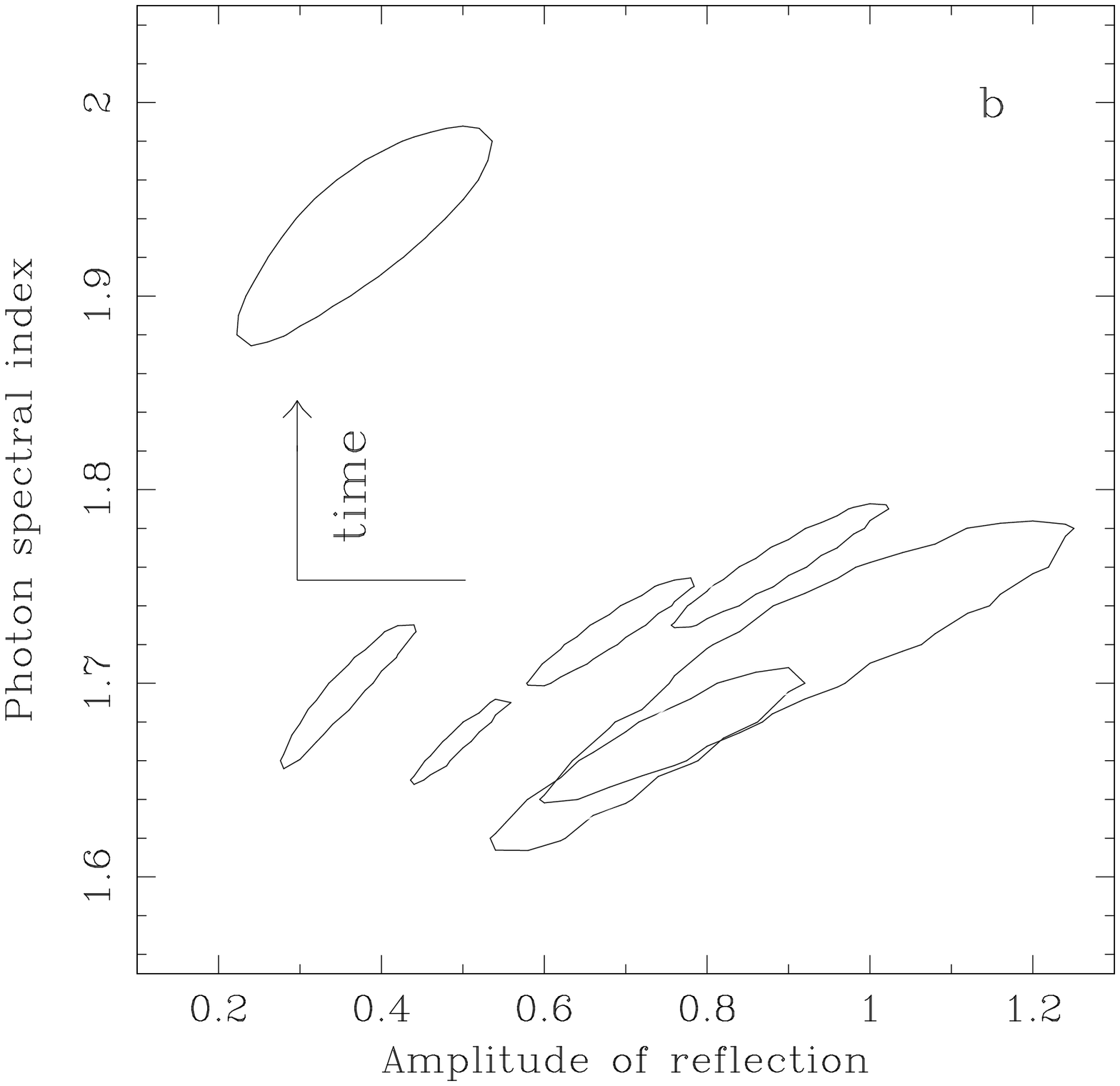,width=2.2in,height=2.2in}
}
\vspace{10pt}
\caption{
Confidence contours as functions of the amplitude of reflection and
photon spectral index for a sequence of spectra; 
(a) -- Nova Muscae 1991, (b) -- GS~2023+338.
}\label{zds:fig2}
\end{figure}
 
Similar results are obtained for another source, GS~2000+25. When a strong,
soft component is present, the power law tail is soft,
the reflector is strongly ionized and
the  relativistic smearing is significant. The disappearance
of the soft component is accompanied by hardening of the power law, 
a drop in ionization and the smearing
effects becoming insignificant (Figure~\ref{zds:fig1}bc).

Thus, in both objects all the observed spectral changes during 
the the HS/LS transition can be explained in a model
where the inner radius of the optically thick flow 
progressively increases: this would trivially explain the decrease in 
solid angle subtended by the disk to the inner X--ray source. Also, as the
luminosity and temperature of the disk declines (since most of the energy is
released in the inner radii) so the ionization state of the disk decreases,
and there are fewer seed photons for Compton scattering giving a harder power 
law, as shown in Figure~\ref{zds:fig2}a.

However, no such clear trend was shown by GS~2023+338. First, the source never
showed ``classical'' soft state spectrum, even though it clearly reached
almost Eddington luminosity \cite{tash96}. During subsequent decline phase, 
the reprocessed component was present in the spectrum, its amplitude 
correlated with the amount of smearing as expected, but this had no 
influence on the spectral index, except that the latter {\it increased\/} 
(i.e.\ spectrum became softer)
sometime between 1 and 6 months after the outburst while the amplitude of
reflection decreased! (Figure~\ref{zds:fig2}b)

Summarizing, the observed transition from high (soft) to low (hard) state 
in Nova Muscae 1991 and GS~2000+25 does seem to
involve a retreat of the optically thick disk, as in the 
EMN model.  However, quantitatively, the inner disk radius 
we infer is much smaller than the values postulated by EMN
in their IS and LS.
Their concept of accretion occurring via an optically thin flow
from very large radii ($10^4\, R_{\rm s}$) for the low state spectra cannot 
be sustained
as such models produce negligible reflected features.  Instead we see 
reflection
at a level which is lower than expected from a complete disk, but is
nonetheless significant. Similar covering fractions for the reflector are seen
in other GBHC hard state spectra, showing that it is a 
general property of these sources \cite{done92,ueda94,zds97}.
While the EMN treatment of the transition radius 
is assumed rather than calculated, nonetheless a
 major facet of their 
model is the sudden switching of {\it all}\/ the accretion flow into an 
optically thin state over a very small range in $\dot{m}$. This is 
clearly inconsistent with the observed behavior of GBHC in general and Nova
Muscae 1991 in particular, where a composite optically thin/optically 
thick flow is required.

Ironically, GS~2023+338 -- the object on which the ADAF model for SXT's
quiescent  states was based \cite{nmcy96} -- does not follow the extended
scenario in a quite spectacular way.

\end{document}